\documentclass[showpacs,twocolumn,aps]{revtex4}
\usepackage{amssymb}
\usepackage{amsmath}
\usepackage{graphicx}
\usepackage{lscape}

\begin{document}

\title{Theoretical analysis for the apparent discrepancy between \={p}p and pp
        data in charged particle forward-backward multiplicity correlations}
\author{Yu-Liang Yan$^{1}$, Bao-Guo Dong$^{1,2}$, Dai-Mei Zhou$^{3}$,
        Xiao-Mei Li$^{1}$,Ben-Hao Sa$^{1,3,4,5}$
        \footnote{Corresponding author: sabh@ciae.ac.cn}}
\address{1) China Institute of Atomic Energy, P. O. Box 275(18), Beijing
102413, China \\
2) Center of Theoretical Physics, National Laboratory of Heavy Ion Collisions,
Lanzhou 730000, China\\
3) Institute of Particle Physics, Huazhong Normal University, Wuhan
430079, China\\
4) CCAST (World Laboratory), P. O. Box 8730 Beijing 100080, China\\
5) Institute of Theoretical Physics, Academy Sciences, Beijing
100080, China}

\begin{abstract}
The strength of charged particle forward-backward multiplicity
correlation in $\bar p+p$ and $p+p$ collisions at $\sqrt s$=200 GeV
is studied by PYTHIA 6.4 and compared to the UA5 and STAR data
correspondingly. It is turned out that a factor of 3-4 apparent
discrepancy between UA5 and STAR data can be attributed to the
differences in detector acceptances and observing bin interval in
both experiments. A mixed event method is introduced and used to
calculate the statistical correlation strength and the dynamical
correlation strengths stemming from the charge conservation, four-
momentum conservation, and decay, respectively. It seems that the
statistical correlation is much larger than dynamical one and the
charge conservation, four-momentum conservation and decay may
account for most part of the dynamical correlation. In addition, we
have also calculated the correlation strength by fitting
the charged particle multiplicity distribution from PYTHIA to the
Negative Binomial Distribution and found that the result
agrees well with the correlation strength calculated by mixed
events.
\end{abstract}


\maketitle

Fluctuations and correlations are important observables
investigating the properties of thermodynamic system and are critical
tools revealing the mechanism of particle production and the
formation of quark-gluon-plasma in relativistic heavy ion collisions
\cite{hwa0,naya}. Several thermodynamic quantities and the produced
particle distributions show varying fluctuation patterns when system
undergoes phase transition. Such as the large energy density
fluctuation is expected in the first order phase transition and a
second order phase transition may relate to a divergence in specific
heat. The event-by-event fluctuation pattern in average transverse
momentum may significantly change around a critical point, etc.

The experimental study of correlation and fluctuation becomes a hot
topic in relativistic heavy ion collisions with the availability of
high multiplicity event-by-event measurements at the CERN-SPS and
BNL-RHIC experiments. There have accumulated an abundant experiment
data \cite{appe,afan,star,star1,star2,phen,phen1,phen2,phob} where
arisen new physics are urgent to be studied.

Recently STAR collaboration has measured the strength of charged particle
forward-backward multiplicity correlation, $b$ (defined later), in
$p+p$ collision at $\sqrt{s}$=200 GeV \cite{star3,star4}. It is 3-4 times
smaller than the one measured by UA5 in $\bar p+p$ collision at the same
energy apparently \cite{ua51983,ua51988}. In this
paper, the PYTHIA 6.4 \cite{soj} is employed to analyze both STAR $p+p$
and the UA5 $\bar{p}+p$ data. It is turned out that the above apparent
discrepancy is because of the differences in detector acceptances
and the interval of pseudo-rapidity bin in both experiments. In
addition, a mixed event method is proposed and used to calculate the
statistical correlation and dynamical correlations stemming from the
charge and four-momentum conservations and the decay of unstable
particles individually. We also fit the particle multiplicity
distribution from PYTHIA to the Negative Binomial Distribution (NBD)
and calculate the strength of charged particle forward-backward
multiplicity correlation which agrees well with the one calculated by
mixed events.

Following Refs \cite{star4,ua51988} the strength of charged particle
forward-backward multiplicity correlation, $b$, is defined
\begin{equation*}
\ b =\frac{\langle n_fn_b\rangle - \langle n_f\rangle \langle
n_b\rangle}{\langle n_f^2\rangle - \langle n_f\rangle^2} =
\frac{cov(n_f,n_b)}{var(n_f)}
\end{equation*}
\begin{equation}
 =\frac{\langle (n_f-\langle n_f\rangle)(n_b-\langle n_b\rangle)\rangle}
  {\langle n_f^2\rangle - \langle n_f\rangle^2},
\label{b}
\end{equation}
where $n_f$ and $n_b$ are, respectively, the number of charged
particles in forward and backward pseudo-rapidity bins ($\Delta
\eta$) defined relatively and symmetrically to a given pseudo-rapidity
$\eta$. The $\langle n_f\rangle$, for instance, refers to the mean value
of $n_f$ and the $cov(n_f,n_b)$ and $var(n_f)$ are, respectively, the
forward-backward multiplicity covariance and forward multiplicity
variance. If there is no correlation between forward and backward
multiplicity, then $\langle n_fn_b\rangle=\langle n_f\rangle \langle
n_b \rangle$ and $b=0$. Thus $b$ is a measure of the strength of
forward-backward multiplicity correlation. As the denominator in Eq.
\ref{b} is positive, if both $n_f$ and $n_b$ are, respectively,
larger or smaller than $\langle n_f\rangle$ and $\langle n_b\rangle$
simultaneously the correlation is positive, negative otherwise.

Although both STAR and UA5 experiments
\cite{star3,star4,ua51983,ua51988} measure the charged particle
multiplicity in Non-Single-Diffractive (NSD) $p+p$ and $\bar p+p$
collisions at $\sqrt s$=200 GeV, the detector acceptances are
different from each other. In UA5 experiment they are $p_T>0$ GeV/c
and $0.0<\vert \eta \vert <4$ \cite{ua51983,ua51988}, but $p_T>0.15$
GeV/c and $0.0<\vert \eta\vert< 1.0$ in STAR \cite{star3,star4}.
Meanwhile, the observed interval of forward- backward pseudo-rapidity
bin is $\Delta\eta$=1.0 in UA5 experiment rather than 0.2 in STAR. We
shall show that those differences are the origin of apparent
discrepancy between STAR and UA5 data in the strength of charged
particle forward-backward multiplicity correlation.

The UA5 data of energy dependence of the correlation strength have
been studied by Dual Parton Model \cite{dpm} and the statistical model
\cite{meng} and the STAR data of forward-backward charged particle
multiplicity covariance have been studied recently in \cite{hwa}. However,
the apparent discrepancy in correlation strength between UA5 and STAR data
is not investigated yet. In this paper PYTHIA 6.4 \cite{soj} is employed
to study that. Since we are not aim to reproduce the experimental data but
to study the physics, we do not adjust the model parameters and default
values are used in all the calculations.

\begin{figure}[htbp]
\includegraphics[height=4.2in,width=3.0in,angle=0]{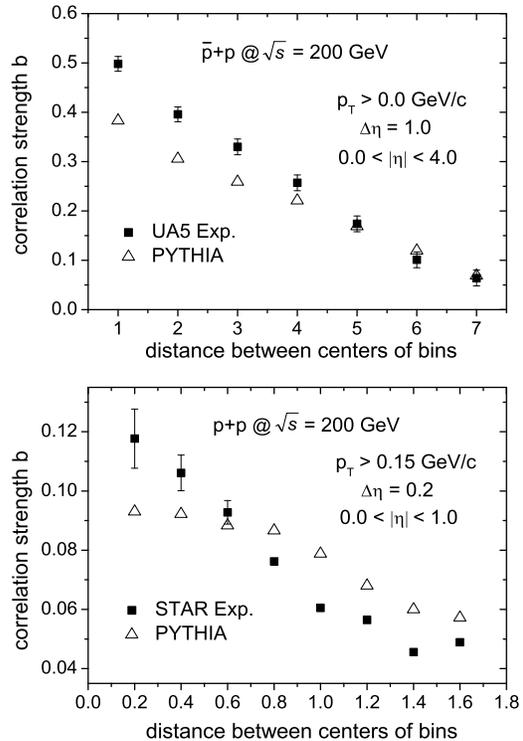}
\caption{The strength $b$ of the charged particle forward-backward
multiplicity correlation in $\bar p+p$ (upper panel) and $p+p$ (lower
panel) collisions at $\sqrt s$=200 GeV. The experimental data are taken
from \cite{ua51988} and \cite{star3}, respectively.}
\label{bexp}
\end{figure}

The comparison of experimental strength of the charged particle
forward-backward multiplicity correlation to the corresponding
theoretical results is given in Figure \ref{bexp} where the upper
panel is for UA5 $\bar p+p$ collision and the lower panel for STAR
$p+p$ collision at $\sqrt s$=200 GeV. One sees in this figure that
the theoretical results are not so far apart from the experimental data
for both the $\bar p+p$ and $p+p$ collisions. The theoretical correlation
strength in $\bar p+p$ collision is also a factor of 3-4 larger than
the one in $p+p$ collision, especially.

\begin{figure}[htbp]
\includegraphics[height=6.3in,width=3.0in,angle=0]{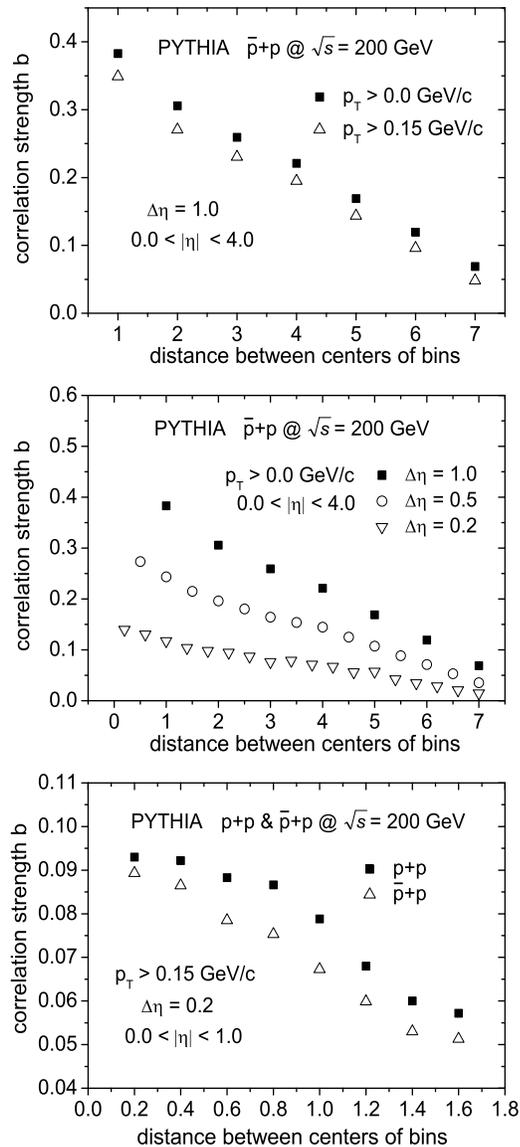}
\caption{Correlation strength $b$: upper panel for $\bar p+p$
collision calculated at different $p_T$ cut; middle panel calculated
at varied intervals of $\eta$ bin; lower panel for $p+p$ and $\bar p+p$
collisions calculated at STAR detector acceptances and pseudo-rapidity
bin interval. The reaction energies are all $\sqrt{s}$=200 GeV.}
\label{bcompare}
\end{figure}

In the upper panel of Figure \ref{bcompare} the full squares are the
PYTHIA results for $\bar p+p$ collision with same detector
acceptances and $\eta$ bin interval as in UA5 experiment, whereas
the open triangles are the PYTHIA results with $p_T>0.15$ instead
of $p_T>0$ GeV/c. The open triangles are monotonously below
the full squares. Middle panel of Figure \ref{bcompare} shows the
PYTHIA results calculated at the same detector acceptances as UA5
but with varied pseudo-rapidity bin intervals: $\Delta\eta$ =1.0
(full squares), 0.5 (open circles), and 0.2 (open triangles-down),
respectively. Here one knows that the correlation strength, $b$, declines
dramatically with the decreasing of pseudo-rapidity bin interval. The
PYTHIA results calculated for both the $p+p$ (full squares) and $\bar
p+p$ (open triangles-up) collisions at STAR detector acceptances and
pseudo-rapidity bin interval are given in lower panel of Figure
\ref{bcompare}. We see in this panel that a factor of 3-4 apparent
discrepancy nearly disappears if both STAR and UA5 experiments are
performed at the same detector acceptances and pseudo-rapidity bin
interval.

It is very hard to separate the statistical and dynamical
correlations (fluctuations) from the measured correlations
(fluctuations) \cite{naya}. We introduce a mixed event method based
on the real (PYTHIA) events (150 thousand events, for instance).
The mixed events are generated one by one according to the real
events. We assume first that the particle multiplicity $N$ in a mixed
event is the same as the corresponding one in real events. However, the
$N$ particles in a mixed event are sampled randomly from the particle
reservoir formed by all particles in the real events. As the
particles in a mixed event are separately and randomly taken from
different real events, there is not any dynamical relevance among them.
Thus the correlation strength, $b$, calculated by the mixed events
is reasonably to be identified as the statistical correlation. Of course,
we can also generate the mixed event with individual constraint, such as
charge conservation, four-momentum conservation, and decay, etc. The
corresponding correlation will be indicated by ``statistical plus
charge dynamical correlations", the ``statistical plus four-momentum
dynamical correlations", and the``statistical plus decay correlations",
etc., respectively.

\begin{figure}[htbp]
\includegraphics[height=4.8in,width=3.5in,angle=0]{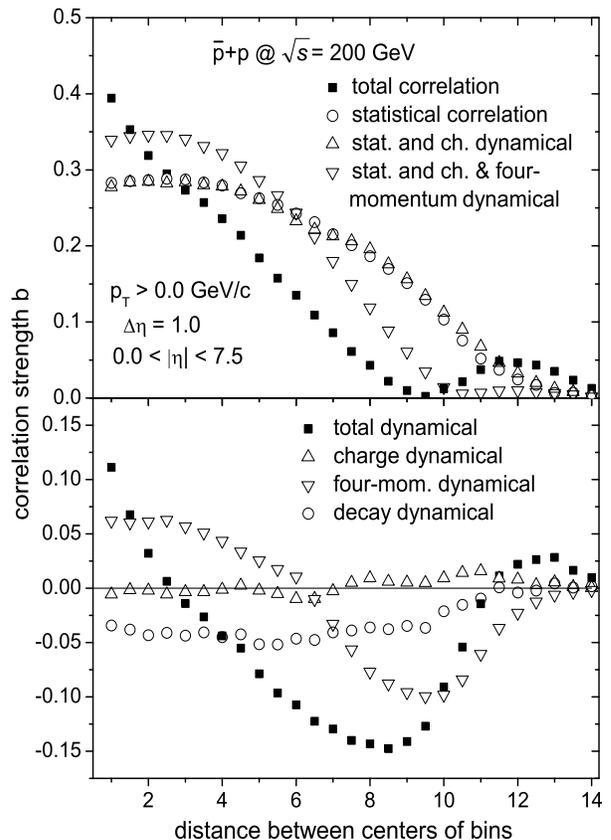}
\caption{Statistical and dynamical correlation strengths. See text
         for detail.}
\label{bmix}
\end{figure}

The strength of charged particle forward-backward multiplicity
correlation is calculated individually from the real
events, mixed events, mixed events with charge conservation, and the
mixed events with charge and four-momentum conservations. They are
given in the upper panel of Figure \ref{bmix} by the full squares
(indicated as total correlation), open circles (statistical
correlation), open triangles-up (statistical plus charge dynamical
correlations) , and the open triangles-down (statistical
plus the charge and four-momentum dynamical correlations),
respectively. In lower panel of Figure \ref{bmix} are given the total
dynamical correlation strength (full squares), the charge dynamical one
(open triangles-up), the four-momentum dynamical one (open
triangles-down), and the decay dynamical correlation (open circles).
The former three are extracted from upper panel by subtracting,
respectively, the statistical correlation from total correlation,
the statistical one from ``statistical plus charge dynamical" one, the
``statistical plus charge dynamical" one from ``statistical plus charge
and four-momentum dynamical" one. It has to mention that in the real
events generated above the decay of unstable hadrons is allowed.
In order to calculate the decay dynamical correlation (open circles in
lower panel of Figure \ref{bmix}), we have first to generate the real
events without decay of unstable hadrons and the corresponding
mixed events. One subtracts the $b$ calculated by mixed events with
(without) decay from the $b$ calculated by real events with (without)
decay, one has the dynamical correlation with (without) decay then. The
decay dynamical correlation in lower panel of Figure \ref{bmix} is just
resulted by subtracting the ``dynamical correlation with decay" from the
one without decay. One knows here that the dynamical correlations stemming
from charge, four-momentum, and the decay may account for the most part
of the total dynamical correlation.

Many experiments have indicated that the particle multiplicity distribution
in hadron-hadron and the nucleus-nucleus collisions is well described by the
Negative Binomial Distribution (NBD) \cite{ua51985,ua51987b,e8021995,phen3}.
For an integer $n$ the NBD reads
\begin{equation}
\ P(n;\mu,k) = \begin{pmatrix}n+k-1\\k-1\end{pmatrix}\frac{
(\mu/k)^n}{(1+\mu/k)^{n+k}},
\label{NBD}
\end{equation}
where $\mu\equiv \langle n \rangle$ is a parameter, $P(n;\mu,k)$ is
normalized in $0\leq n \leq\infty$, and $k$ is another parameter responsible
for the shape of the distribution. If $k$ is a real the binomial coefficient
in Eq. \ref{NBD} is $k(k+1)\cdots(k+n-1)/n!$. In NBD the variance
($\sigma^2$) and mean ($\mu$) is related to $k$ by
\begin{equation}
\ \sigma^{2} = \mu +\frac{\mu^2}{k}.
\label{b1}
\end{equation}
An important property of NBD is that if particle multiplicity $n$ is
NBD in whole phase space and the particle has unified probability,
$p$, in a partial phase space (such as in a $\eta$ bin here) then
particle multiplicity distribution in this partial phase space is
also NBD with same parameter $k$ and the mean is equal to $\mu p$
\cite{adc}. That is obviously based on the assumption that the
particles are independent with each other, i.e. there is no
dynamical correlation among them.

\begin{figure}[htbp]
\includegraphics[height=2.3in,width=3.3in,angle=0]{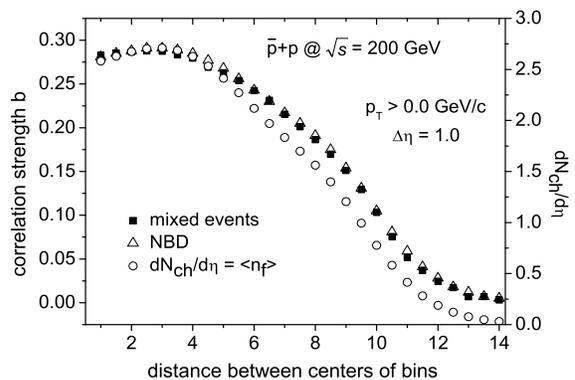}
\caption{Charged particle forward-backward multiplicity statistical
correlation strength $b$ and the charged particle pseudo-rapidity
distribution in $\bar p+p$ collision at $\sqrt{s}$=200 GeV.}
\label{bk}
\end{figure}

Since forward and backward pseudo-rapidity bins are symmetry
relative to the investigated pseudo-rapidity ($\eta$) and have same
width so $\langle n_f\rangle=\langle n_b\rangle$ and
$var(n_f)=var(n_b)$ in NBD. Using the statistic formula \cite{myer}
\begin{equation}
\ var(X\pm Y)=var(X)+var(Y)\pm 2cov(X,Y)
\end{equation}
the $b$ can be written as
\begin{equation}
\ b =\frac{cov(n_f,n_b)}{var(n_f)}=\frac{var(n_f+n_b)-2var(n_f)}{2var(n_f)}.
\label{b2}
\end{equation}
Substitute Eq. \ref{b1} into Eq. \ref{b2} one has
\begin{equation}
\ b = \frac{\langle n_f\rangle}{\langle n_f\rangle+k}.
\label{b3}
\end{equation}
We know that the NBD becomes a Poisson distribution in the limit
$k\rightarrow\infty$, so the correlation strength, $b$, is zero in
Poisson distribution. If the charged particle multiplicity
distribution in real events with decay assumption is fitted by NBD,
the parameter $k$ is obtained. As the real events are generated in
NSD (Non-Single-Diffractive) indeed, the charged particle
multiplicity distribution is not perfect NBD, therefore the above
fit is not so sensitive to the $k$ values within 6-7. If NBD with
$k$=6.6 is assumed for the charged particle multiplicity
distribution, the corresponding $b$ can be calculated by Eq.
\ref{b3} because $\langle n_f\rangle$ can be approximated by
$dN_{ch}$/$d\eta$ in real event. Those $b$ are shown in Fig.
\ref{bk} by open triangles. In this figure the full squares are
calculated by the mixed events with decay assumption (i.e. the open
circles in upper panel of Fig. \ref{bmix}) and the open circles are
the charged particle pseudo-rapidity distribution in real events
with decay (in drawing $dN_{ch}$/$d \eta$ the abscissa is identified
as $\eta$ and scaled by 2). The results of NBD agree well with the
results calculated by the mixed events with decay assumption, it
proves again that we are reasonable identifying the $b$ calculated
by mixed events as the statistical correlation strength. Comparing
the full squares and open triangles to the open circles one knows
that the statistical correlation strength may have shape similar to
the charged particle pseudo-rapidity distribution.

Recently, the forward-backward multiplicity covariance in $p+p$
collision at $\sqrt s$=200 GeV has been studied in Ref. \cite{hwa}.
They assumed the back-to-back partonic scattering is the origin of
hadronic correlation, related that partonic scattering angles to a
Gaussian like hadronization function, and derived the forward-backward
multiplicity covariance. Without more dynamical input their results
are well comparing with STAR data \cite{star3}. Therefore they conclude
that the correlation length might have no fundamental significance. We
plan to investigate the partonic origin of forward-backward multiplicity
correlation by transport model in next study.

In summary, we have calculated the strength of charged particle
forward-backward multiplicity correlation in $\bar p+p$ and $p+p$
collisions at $\sqrt s$=200 GeV by PYTHIA 6.4 \cite{soj} and
compared with UA5 data \cite{ua51988} and STAR data \cite{star3},
respectively. It is turned out that a factor of 3-4 apparent
discrepancy between UA5 and STAR data can be attributed to the
differences in detector acceptances and the interval of observed
$\eta$ bin in both experiments. A mixed event method is introduced and
used to calculate the statistical correlation strength and the
individual dynamical correlations stemming from charge conservation,
four-momentum conservation, and the decay. It seems that the
statistical correlation is much larger than the dynamical one, and
the charge, four-momentum, and decay may account for the main part
of the dynamical correlation. The NBD $b$ results agree well with
the ones calculated by mixed events proves again that one is
reasonable to identify the correlation in mixed events as a
statistical correlation.

The financial support from NSFC (10635020, 10705012, and 10605040)
in China is acknowledged.

\end{document}